\documentstyle[epsfig,amsfonts]{article}
\def \arctanh{\mathop{\rm arctanh}\nolimits}

\newcommand{\beq}{\begin{eqnarray}}
\newcommand{\eeq}{\end{eqnarray}}
\newcommand{\eqn}{\begin{equation}}
\newcommand{\een}{\end{equation}}
\begin{document}
\title{N=2 Supersymmetric Kinks and real algebraic curves}
\author{\dag A. Alonso Izquierdo, \dag M.A.
Gonz\'alez Le\'on and \ddag J. Mateos Guilarte \\ {\it \dag
Departamento de Matem\'atica Aplicada, \ddag Departamento de
F\'{\i}sica} \\ {\it Universidad de Salamanca, SPAIN}}
\date{}
\maketitle
\begin{abstract}
The kinks of the (1+1)-dimensional Wess-Zumino model with
polynomic superpotential are investigated and shown to be related
to real algebraic curves.
\end{abstract}



\section{}
The dimensional reduction of the (3+1)-dimensional Wess-Zumino
model, produces an interesting (1+1)-dimensional Bose-Fermi
system; this field theory enjoys N=2 extended supersymmetry
provided that the interactions are introduced via a real harmonic
superpotential, see \cite{Sh}. In a recent paper \cite{Gi} Gibbons
and Townsend have shown the existence of domain-wall intersections
in the (3+1)D WZ model, the authors relying on the supersymmetry
algebra of the (2+1)D dimensional reduction of the system.
Although the domain-wall junctions are two-dimensional structures,
their properties are reminiscent of the one-dimensional kinks from
which they are made. In this letter we shall thus describe the
kinks of the underlying (1+1)-dimensional system.

The basic fields of the theory are:
\begin{itemize}
\item  Two real bosonic fields, $\phi^a(x^\mu)$, $a=1,2$ that can
be assembled in the complex field: $\phi(x^\mu)=\phi^1(x^\mu)+i
\phi^2 (x^\mu)\in {\rm Maps}({\Bbb R}^{1,1},{\Bbb C})$.
$x^\mu=(x^0,x^1)$ are local coordinates in the ${\Bbb R}^{1,1}$
Minkowski space, where we choose the metric $g^{\mu \nu},
g^{00}=-g^{11}=1, g^{12}=g^{21}=0$.

\item Two Majorana spinor fields $\psi^a(x^\mu)$, $a=1,2$. We work
in a Majorana representation of the Clifford algebra
$\{\gamma^\mu, \gamma^\nu\}=2 g^{\mu \nu}$,
\[
\gamma^0=\sigma^2, \hspace{1cm} \gamma^1=i \sigma^1, \hspace{1cm}
\gamma^5=\gamma^0 \gamma^1=\sigma^3
\]
where $\sigma^1$, $\sigma^2$, $\sigma^3$ are the Pauli matrices,
such that ${\psi^a}^{*}= \psi^a$. We also define the adjoint
spinor as $\bar{\psi}(x^\mu)=\psi^{t} (x^\mu) \gamma^0$ and
consider Majorana-Weyl spinors: $\psi^a_{\pm}(x^\mu)=\frac{1\pm
\gamma^5}{2} \psi^a(x^\mu)$ with only one non-zero component.
\end{itemize}

\noindent Interactions are introduced through the holomorphic
superpotential: $W(\phi)=\frac{1}{2} \left( W^1(\phi^1,\phi^2)+i W^2(\phi^1,\phi^2)
\right)$. One could in principle start from the supercharges:
\[
{\hat Q}_{\pm}^{BC}=\int dx^1 \sum_{a,b} \left[f^B\right]^{ab}
\left[ (\partial_0 \phi^a \mp
\partial_1 \phi^a) \psi_{\pm}^b \pm \sum_c \left[f^C\right]^{bc}
 \frac{\partial W^C}{\partial
\phi^c} \psi^a_{\mp} \right]
\]
where $W^B$, $B=1,2$, are respectively the real part if $B=1$ and
the imaginary part if $B=2$ of $W(\phi)$ and $\left[f^B\right]$ is
either the identity or the complex structure endomorphism in
${\Bbb R}^2$ \cite{Fr}:
\[
\left[f^{B=1}\right]=\left( \begin{array}{cc} 1 & 0 \\ 0 & 1
\end{array} \right) \hspace{1cm} \left[f^{B=2}\right]=\left(
\begin{array}{cc} 0 & 1 \\ -1 & 0 \end{array} \right).
\]
Nevertheless, the Cauchy-Riemann equations:
\begin{equation}
\frac{\partial W^1}{\partial \phi^1} = \frac{\partial
W^2}{\partial \phi^2} \hspace{2cm} \frac{\partial W^1}{\partial
\phi^2} =- \frac{\partial W^2}{\partial \phi^1},
\end{equation}
tell us that the theory is fully described by choosing either
$W^1$ or $W^2$. We thus set $W^C=W^1$ and find the basic SUSY
charges to be ${\hat Q}_{\pm}^{B1}=Q_{\pm}^B$:
\begin{equation}
Q_{\pm}^B=\int dx^1 \sum_{a,b} \left[f^B\right]^{ab}  \left[
(\partial_0 \phi^a \mp
\partial_1 \phi^a) \psi_{\pm}^b \pm
 \frac{\partial W^1}{\partial
\phi^b} \psi^a_{\mp} \right] \label{eq:ch}
\end{equation}
 From the canonical quantization rules
\begin{equation}
[\phi^a(x_1),\dot{\phi}^b(y_1) ]=i \delta^{ab} \delta(x_1-y_1)=\{\psi^a_{\pm}(x_1),i \psi^b_{\pm}(y_1)\}
\end{equation}
one checks that the $N=2$ extended supersymmetric algebra
\begin{equation}
\{ Q_{\pm}^B, Q_{\pm}^C \}=2 \delta^{BC} P_{\mp} \hspace{2cm}
\{Q_+^B,Q_-^C\}= -(-1)^B( \delta^{BC}2T+ \epsilon^{BC} 2\tilde{T})
\label{eq:alge}
\end{equation}
is closed by the four generators $Q_{\pm}^B$, defined in
(\ref{eq:ch}). Here
\begin{eqnarray*}
P_{\pm}&=&\frac{1}{2} \int d x^1 \sum_a \left[(\partial_0
\phi^a\pm
\partial_1 \phi^a)(\partial_0 \phi^a \pm \partial_1 \phi^a) \pm 2
i \psi_{\mp}^a \partial_1 \psi^a_\mp \right] + \\ &+& \frac{1}{2}
\int d x^1 \sum_a \left[ \frac{\partial W^1}{\partial \phi^a}
\frac{\partial W^1}{\partial \phi^a} -2 i \sum_b \frac{\partial^2
W^1}{\partial \phi^a \partial \phi^b} \psi_+^a \psi_-^b \right]
\end{eqnarray*}
are the light-cone momenta and
\[
T=\int dx^1 \left[  \frac{\partial W^1}{\partial \phi^1}
\frac{\partial \phi^1}{\partial x^1}+ \frac{\partial W^1}{\partial
\phi^2} \frac{\partial \phi^2}{\partial x^1}\right]=\int d
W^1=W^1(\infty)-W^1(-\infty)
\]
\[
\tilde{T}=\int dx^1 \left[  \frac{\partial W^1}{\partial \phi^1}
\frac{\partial \phi^2}{\partial x^1}-\frac{\partial W^1}{\partial
\phi^2} \frac{\partial \phi^1}{\partial x^1}\right]=\int * d
W^1=W^2(\infty)-W^2(-\infty)
\]
the central extensions.

\section{}

From the SUSY algebra one deduces,
\[
2P_0=2 |T|+( Q_+^B \pm (-1)^B Q_-^B )^2=2 |{\tilde T}|+( Q_\pm^B
\mp (-1)^B \epsilon^{BC} Q_\mp^C)^2 ,
\]
see \cite{OW}. We thus define the charge operators on zero
momentum states:
\begin{eqnarray*}
\tilde{Q}_\pm^1&=&Q_+^1 \pm Q_-^1=  -\int dx^1 \sum_a \left(
\partial_1 \phi^a \pm \frac{\partial W^1}{\partial \phi^a} \right)
\left( \psi_+^a \mp \psi_-^a\right) \\
\tilde{Q}_\pm^2&=&Q^1_\pm \pm Q^2_\mp=  \mp \int dx^1 \sum_a \left(
\partial_1 \phi^a \pm \sum_b \left[f^2\right]^{ab} \frac{\partial
W^1}{\partial \phi^b} \right) \left( \psi_\pm^a \mp \sum_c
\left[f^2\right]^{ac} \psi_\mp^c\right)
\end{eqnarray*}

Spatially extended coherent states built from the solutions of any
of the two systems of first order equations, \cite{Ba}:
\begin{equation}
\frac{d \phi^1}{d x^1}=\pm\frac{\partial W^1}{\partial \phi^1}
\hspace{1cm} \frac{d \phi^2}{d x^1}=\pm\frac{\partial
W^1}{\partial \phi^2} \label{eq:ecudif1}
\end{equation}
\begin{equation}
\frac{d \phi^1}{d x^1}=\pm\frac{\partial W^1}{\partial \phi^2}
\hspace{1cm} \frac{d \phi^2}{d x^1}=\mp\frac{\partial
W^1}{\partial \phi^1} \label{eq:ecudif2}
\end{equation}
have minimum energy because they are respectively annihilated by
${\tilde Q}_{\pm}^1$ (system (\ref{eq:ecudif1})) and ${\tilde
Q}_{\pm}^2$ (system (\ref{eq:ecudif2})) .

The flow in ${\Bbb R}^2 \simeq {\Bbb C}$ of the solutions of
(\ref{eq:ecudif1}) is given by:
\[
\frac{d \phi^2}{d \phi^1}=\frac{\partial W^1}{\partial \phi^2}
\left( \frac{\partial W^1}{\partial \phi^1} \right)^{-1}
\hspace{0.5cm} \equiv \hspace{0.5cm} \frac{\partial W^1}{\partial
\phi^2} d\phi^1-\frac{\partial W^1}{\partial \phi^1} d\phi^2 =
dW^2=0
\]
If $W(\phi)$ is polynomic in $\phi$, the solutions of
(\ref{eq:ecudif1}) live on the real algebraic curves determined by
the equation:
\begin{equation}
 W^2(\phi^1,\phi^2)= \gamma_\perp \label{eq:curvas1}
\end{equation}
where $\gamma_\perp$ is a real constant. Simili modo, the solution
flow of (\ref{eq:ecudif2}) in ${\Bbb C}$,
\[
\frac{d \phi^2}{d \phi^1}=-\frac{\partial W^1}{\partial \phi^1}
\left( \frac{\partial W^1}{\partial \phi^2} \right)^{-1}
\hspace{0.5cm} \equiv \hspace{0.5cm} \frac{\partial W^1}{\partial
\phi^1} d\phi^1+\frac{\partial W^1}{\partial \phi^2} d\phi^2 =d
W^1=0
\]
runs on the real algebraic curves:
\begin{equation}
W^1(\phi^1,\phi^2)=\gamma \label{eq:curvas2}
\end{equation}
where $\gamma$ is another real constant. There are two
observations: (I) Solutions of system (\ref{eq:ecudif1}) live on
curves for which $W^2=constant$  and solutions of
(\ref{eq:ecudif2}) have support on curves for which
$W^1=constant$. (II) The curves that support the solutions of
(\ref{eq:ecudif1}) are orthogonal to the curves related to the
solutions of (\ref{eq:ecudif2}).

Assume that $W(\phi)$ has a discrete set of extrema, forming the
vacuum orbit of the system: $\left. \frac{\partial W}{\partial
\phi} \right|_{v^{(i)}}=0 $, $i=1,2,...,n$. Kinks are solutions of
(\ref{eq:ecudif1}) and/or (\ref{eq:ecudif2}) such that they tend
to $v^{(i_\pm)}$ when $x_1$ reaches $\pm \infty$. $v^{(i_+)}$ and
$v^{(i_-)}$ thus belong  either to curves (\ref{eq:curvas1}) or
(\ref{eq:curvas2}), and this fixes the values of $\gamma$ or
$\gamma_\perp$ for which the real algebraic curves support kinks.
In Reference \cite{Va} a general proof based in singularity theory
of the existence of these soliton solutions, that counts its
number, is achieved. The energies of the states grown from kinks
are $P_0=| T|=\left| W^1(v^{(i_+)}) -W^1(v^{(i_-)}) \right|$ for
solutions of (\ref{eq:ecudif1}) and $P_0=|\tilde{T}| = \left|
W^2(v^{(i_+)})-W^2(v^{(i_-)}) \right|$ for solutions of
(\ref{eq:ecudif2}). The kink form factor is obtained from a
quadrature: one replaces either (\ref{eq:curvas1}) or
(\ref{eq:curvas2}) in the first equation of (\ref{eq:ecudif1}) or
(\ref{eq:ecudif2}) and integrates.

Therefore, the fermionic charges ${\tilde Q}_{\pm}^1$ and ${\tilde
Q}_{\pm}^2$ are annihilated on coherent states $\left| K_{\pm}^1
\right>$ and $\left| K_{\pm}^2\right>$ that correspond to the
tensor product of the quantum antikink/kink, living respectively
on curves $W^2=constant$ and $W^1=constant$, with its
supersymmetric partners (the translational mode times a constant
spinor). We find
\begin{eqnarray*}
{\tilde Q}^1_{\pm}\left| K_{\pm}^1 \right>&=&\int dx_1 \sum_a
\left[\partial_1\phi_{K^1_{\pm}}^a \pm \left. \frac{\partial
W^1}{\partial\phi^a}\right| \rule{0in}{.3in}_{\phi_{K^1_{\pm}}^a}\right]\partial_1\phi_{K^1_{\pm}}^a
\left(\begin{array}{c} 1 \\ {\mp}1 \end{array} \right) \left|
K_{\pm}^1 \right>=0\\
{\tilde Q}^2_+\left| K_+^2 \right>&=&\int dx_1 \sum_a
\left[\partial_1\phi_{K^2_+}^a + \sum_b \epsilon^{ab} \left.\frac{\partial
W^1}{\partial\phi^b}\right|\rule{0in}{.3in}_{\phi_{K^2_+}^a}\right] \left(\begin{array}{c}
\partial_1\phi_{K^2_+}^a
\\ -\sum_c \epsilon^{ac} \partial_1\phi_{K^2_+}^c \end{array} \right) \left| K_+^2 \right>=0 \\
{\tilde Q}^2_-\left| K_-^2 \right>&=&\int dx_1 \sum_a
\left[\partial_1\phi_{K^2_-}^a - \sum_b \epsilon^{ab} \left.\frac{\partial
W^1}{\partial\phi^b}\right|\rule{0in}{.3in}_{\phi_{K^2_-}^a}\right] \left(\begin{array}{c}
\sum_c \epsilon^{ac} \partial_1\phi_{K^2_-}^c
\\\partial_1\phi_{K^2_-}^a  \end{array} \right) \left| K_+^2 \right>=0
\end{eqnarray*}
on solutions of (7) and/or (8); the SUSY kinks are thus
$\frac{1}{4}$-BPS states. The energy of these states does not
receive quantum corrections \cite{Sh}, because $N=2$ supersymmetry
forbids any anomaly in the central charges.

\section{}
We focus on the case in which the potential is:
\[
U(\phi)=\frac{1}{2} \sum_a  \frac{\partial W^1}{\partial
\phi^a} \frac{\partial W^1}{\partial \phi^a} =\frac{1}{2}\left(1- 2
(\phi_1^2+ \phi_2^2)^{\frac{n-1}{2}} \cos \left[ (n-1) \arctan
\frac{\phi^2}{\phi^1}\right]  + (\phi_1^2 +\phi_2^2)^{n-1}\right)
\]
see \cite{Gi} and \cite{To}. In polar variables in the ${\Bbb
R}^2$ internal space,

\[
\rho(x^\mu)=+\sqrt{[\phi^1(x^\mu)]^2 +[\phi^2(x^\mu)]^2 }, \hspace{1cm} \chi(x^\mu)=\arctan \frac{\phi^2(x^\mu)}{\phi^1(x^\mu)}
\]
 the potential reads:
\begin{equation}
U(\rho,\chi)=\frac{1}{2}\left(1-2 \rho^{n-1} \cos (n-1)
\chi+\rho^{2(n-1)}\right) \label{eq:pot}
\end{equation}
There is symmetry under the $D_{2(n-1)}\equiv {\Bbb Z}_2 \times
{\Bbb Z}_{n-1}$ dihedral group: $\chi' =-\chi$,
$\chi'=\chi+\frac{2 \pi j}{n-1}$, $j=0,1,2,...,n-2$. In Cartesian
coordinates, these transformations form the $D_{2(n-1)}$ sub-group
of $O(2)$ given by:
\begin{itemize}
\item[(1)] $\phi_2'=-\phi_2, \hspace{0.4cm} \phi_1'=\phi_1$

\item[(2)] $ {\phi^1}'= \cos \displaystyle\frac{2 \pi j}{n-1} \phi^1- \sin \displaystyle\frac{2 \pi j}{n-1} \phi^2, \hspace{0.4cm} {\phi^2}'= \sin \displaystyle\frac{2 \pi j}{n-1} \phi^1+ \cos \displaystyle\frac{2 \pi j}{n-1} \phi^2 $
\end{itemize}

The vacuum orbit is the set of $(n-1)$-roots of unity:
\begin{equation}
{\cal M}=\left\{  v^{(k)}=e^{i \frac{2 \pi k}{n-1}}  \right\} =
\frac{D_{2(n-1)}}{{\Bbb Z}_2} = {\Bbb Z}_{n-1} .
\end{equation}
When the $v^{(k)}$ vacuum is chosen to quantize the theory, the
symmetry under the $D_{2(n-1)}$ group is spontaneously broken to
the ${\Bbb Z}_2$ sub-group generated by $\chi'=-\chi-\frac{2 \pi
k}{n-1}$; this transformation leaves a fixed point, $v^{(k)}$, if
$n$ is even and two fixed points, $v^{(k)}$ and
$v^{(k+\frac{n-1}{2})}$, if $n$ is odd.

The ${\Bbb Z}_{n-1}$-symmetry allows for the existence of $(n-1)$
harmonic superpotentials that are equivalent:
$W^{(j)}(\phi)=\frac{1}{2}\left[ \phi^{(j)}
-\frac{(\phi^{(j)})^n}{n} \right]$, $\phi^{(j)}=e^{i \frac{2 \pi
j}{n-1}}\phi$, all of them leading to the same potential $U$.
Thus:
\[
W^{(j)1}=\rho \cos \chi(j)-
\frac{1}{n} \rho^n \cos n \chi(j) \hspace{1.2cm} W^{(j)2}=\rho \sin \chi(j)- \frac{1}{n} \rho^n \sin  n \chi(j)
\]
where $\chi(j)=\chi+\frac{2 \pi j}{n-1}$. There is room for
closing the $N=2$ supersymmetry algebra (\ref{eq:alge}) in $n-1$
equivalent forms: define the $n-1$ equivalent sets of SUSY
charges:
\[
{Q_{\pm}^{(j)}}^B =\int d x^1 \sum_{a,b}
\left[\left[f^B\right]^{ab} (\partial_0 \phi^{(j)a} \mp
\partial_1 \phi^{(j)a}) \psi_{\pm}^{(j)b} \pm
\frac{\partial W^{(j)1}}{\partial \phi^{(j)a}} \psi_{\mp}^{(j)b}
\right],
\]
also in terms  of the "rotated" fermionic fields
$\psi_{\pm}^{(j)a}$ , and the corresponding central charges
$T^{(j)}$ and ${\tilde T}^{(j)}$. Observe that the $N=2$
supersymmetry is unbroken, while the choice of vacuum that
spontaneously breaks the ${\Bbb Z}_{n-1}$ symmetry does not affect
the physics, which is the same for different values of $j$.

The $j$ pairs of first-order systems of equations:
\begin{equation}
\displaystyle\frac{d \rho}{d x_1}=  \sin
\chi(j)-  \rho^{n-1} \sin
n \chi(j) \hspace{1cm} \rho^2
\displaystyle\frac{d \chi(j)}{d x_1}=  \rho \cos
\chi(j)-  \rho^{n} \cos
n \chi(j)
\label{eq:rota2}
\end{equation}
\begin{equation}
\displaystyle\frac{d \rho}{d x_1}=  \cos
\chi(j)-  \rho^{n-1} \cos
n \chi(j)\hspace{1cm} \rho^2
\displaystyle\frac{d \chi(j)}{d x_1}= - \rho \sin
\chi(j)+  \rho^{n} \sin
n \chi(j)
\label{eq:rota1}
\end{equation}
correspond to (\ref{eq:ecudif1}) and (\ref{eq:ecudif2}) for this
particular case. The solutions lie respectively on the algebraic
curves
\begin{eqnarray}
\rho \sin \chi(j)-\frac{1}{n} \rho^n \sin  n\chi(j)&=&
\gamma_\perp \label{eq:tra1} \\ \rho \cos \chi(j)-\frac{1}{n}
\rho^n \cos n\chi(j)&=& \gamma \label{eq:tra2}
\end{eqnarray}
which form two families of orthogonal lines in ${\Bbb R}^2$. In the family of curves (\ref{eq:tra1}) there are kinks joining the vacua $v^{(k)}$ and $v^{(k')}$ if and only if:
\begin{equation}
\sin \frac{2 \pi (k+j)}{n-1} - \frac{1}{n} \sin \frac{2 \pi (k+j)
n}{n-1}=\sin \frac{2 \pi (k'+j)}{n-1} - \frac{1}{n} \sin \frac{2
\pi (k'+j) n}{n-1}=\gamma_\perp^K \label{eq:condini2}
\end{equation}
This fixes the value of $\gamma_\perp=\gamma_\perp^K$ for which
the algebraic curve supports a topological kink. Simili modo,
\begin{equation}
\cos \frac{2 \pi (k+j)}{n-1} - \frac{1}{n} \cos \frac{2 \pi (k+j)
n}{n-1}=\cos \frac{2 \pi (k'+j)}{n-1} - \frac{1}{n} \cos \frac{2
\pi (k'+j) n}{n-1}=\gamma^K \label{eq:condini1}
\end{equation}
is the value of the constant if the kink belong to the orthogonal
family (\ref{eq:tra2}). Solutions of (\ref{eq:condini2}) and/or
(\ref{eq:condini1}) exist, respectively, if and only if
\begin{equation}
2( k+k'+2 j)= n-1 \, {\bf mod} \, 2(n-1)
\end{equation}
and/or
\begin{equation}
k+k'+2 j=0 \, {\bf mod} \, n-1
\end{equation}

Given the kink curves, the kink form factors are obtained in the
following way:

\noindent One solves for $\chi$ in (\ref{eq:tra1}) or
(\ref{eq:tra2}),
\begin{equation}
\chi+\frac{2\pi j}{n-1}=h(\gamma^K, \rho) \hspace{1.5cm}, \hspace{1.5cm}
\chi+\frac{2\pi j}{n-1}=h_\perp(\gamma_\perp^K,\rho)
\end{equation}
and plugs these expressions into the first equation of
(\ref{eq:rota1}) or (\ref{eq:rota2}),
\[
\frac{d \rho}{d x_1}= \sin h(\gamma^K,\rho)-  \rho^{n-1} \sin [n
h(\gamma^K,\rho)]\hspace{0.4cm}, \hspace{0.4cm} \frac{d \rho}{d x_1}= \cos
h_\perp(\gamma_\perp^K,\rho)-  \rho^{n-1} \cos [n
h_\perp(\gamma_\perp^K,\rho)]
\]
which are immediately integrated by quadratures: if $a$ is an
integration constant
\begin{eqnarray}
\int \frac{d \rho}{\sin h(\gamma^K,\rho)-\rho^{n-1} \sin [n
h(\gamma^K,\rho)]} &=& (x_1+a) \\ \int \frac{d \rho}{\cos
h_\perp(\gamma_\perp^K,\rho)-\rho^{n-1} \cos [n h_\perp(\gamma_\perp^K,\rho)]} &=& (x_1+a)
\end{eqnarray}

\section{}
We first consider  the lower odd cases, only for $W^{(j=0)}$. The
other kinks are obtained by application of a ${\Bbb Z}_{n-1}$
rotation.
\begin{itemize}
\item $n=3$:
\begin{itemize}
\item {\bf Superpotential:} $W(\phi)=\frac{1}{2}\left(\phi-\frac{\phi^3}{3}\right)$
\[
W^1=\phi_1-\frac{\phi_1^3}{3}+\phi_1 \phi_2^2
\hspace{1cm} W^2=\phi_2-\phi_1^2 \phi_2+\frac{\phi_2^3}{3}
\]
\item {\bf Potential:} $U(\phi_1,\phi_2)=\frac{1}{2} [(\phi \phi^{*}-1)^2+4 \phi_2^2] $
\item {\bf Vacuum orbit:} ${\cal M}=\frac{D_2}{{\Bbb Z}_2}=\{v^0=1, v^1=-1\} $
\item {\bf Real algebraic curves:}
\[
\phi_1-\frac{\phi_1^3}{3}+\phi_1 \phi_2^2= \gamma ; \hspace{1cm} \phi_2-\phi_1^2 \phi_2+\frac{\phi_2^3}{3}= \gamma_\perp
\]
\item {\bf Kink curve:} $\gamma_\perp=0 (\equiv W^2=0)$, tantamount to $\phi_2=0$.
\item {\bf Kink form factor:}

a) Solutions of $\frac{d \phi_1}{d x_1}=\pm (1-\phi_1^2)$ on $\phi^2=0$: $\phi_1^{K^1_{\mp}}(x_1)=\pm \tanh (x_1+a)$

\item {\bf Kink energy:} $P_0[\phi^{K_{\pm}^1}]=|T|= \left| W^1(v^0)-W^1(v^1) \right|=\frac{4}{3}$
\item {\bf Conserved SUSY charge:} ${\tilde Q}_{\pm}^1 \left| K_{\pm}^1\right>=0$
\end{itemize}

\item $n=5$:
\begin{itemize}
\item {\bf Superpotential:}  $W(\phi)=\frac{1}{2}\left(\phi-\frac{\phi^5}{5}\right)$
\[
W^1=\phi_1 \left(1-\frac{\phi_1^4}{5}+2 \phi_1^2 \phi_2^2- \phi_2^4\right) \hspace{1cm} W^2= \phi_2\left(1-\phi_1^4+2
\phi_1^2 \phi_2^2-\frac{\phi_2^4}{5}\right)
\]
\item {\bf Potential:} $U(\phi_1,\phi_2)=\frac{1}{2} [(\phi \phi^{*}+1)^2-4 \phi_1^2] [(\phi \phi^{*}+1)^2-4 \phi_2^2] $
\item {\bf Vacuum orbit:} ${\cal M}=\frac{D_4}{{\Bbb Z}_2}=\{v^0=1, v^1=i, v^2=-1, v^3=-i\} $
\item {\bf Real algebraic curves:}
\[
\phi_1\left(1 -\frac{\phi_1^4}{5}+2 \phi_1^2 \phi_2^2-\phi_2^4\right)=\gamma;\hspace{0.4cm} \phi_2 \left(1-\phi_1^4+2 \phi_1^2 \phi_2^2-\frac{\phi_2^4}{5} \right) =\gamma_\perp
\]
\item {\bf Kink curves:} a) $\gamma_\perp=0 \equiv \phi_2=0$, b) $\gamma=0 \equiv \phi_1=0$.
\item {\bf Kink form factor:}
\begin{itemize}
\item[a)] Solutions of $\pm \frac{d \phi_1}{d x_1}=1-\phi_1^4$
on $\phi_2=0$: $\arctan \phi_1^{K_\mp^1}+\arctanh \phi_1^{K_\mp^1}= \pm 2
(x_1+a)$

\item[b)] Solutions of $\pm \frac{d \phi_2}{d x_1}=1-\phi_2^4$
on $\phi_1=0$ : $\arctan \phi_2^{K_\mp^2}+\arctanh \phi_2^{K_\mp^2}=\pm 2
(x_1+a)$
\end{itemize}

\item {\bf Kink energies:} (a) $P_0[\phi^{K_{\pm}^1}]=|T|= \left| W^1(v^0)-W^1(v^2) \right|=\frac{8}{5}$

\hspace*{2.6cm} (b) $P_0[\phi^{K_{\pm}^2}]=|{\tilde T}|= \left|
W^2(v^1)-W^2(v^3) \right|=\frac{8}{5}$

\item {\bf Conserved SUSY charges:} (a) ${\tilde Q}_{\pm}^1 \left|
K_{\pm}^1\right>=0$; (b) ${\tilde Q}_{\pm}^2\left|
K_{\pm}^2\right>=0$
\end{itemize}
\item $n=7$:
\begin{itemize}
\item {\bf Superpotential:} $W[\phi]=\frac{1}{2}\left(\phi-\frac{\phi^7}{7}\right)$
\[
W^1=\phi_1-\frac{\phi_1^7}{7}+3 \phi_1^5 \phi_2^2- 5 \phi_1^3
\phi_2^4+ \phi_1 \phi_2^6 \hspace{1cm} W^2= \phi_2 - \phi_1^6
\phi_2 +5 \phi_1^4 \phi_2^3 -3\phi_1^2
\phi_2^5+\frac{\phi_2^7}{7}
\]
\item {\bf Potential:} $U(\phi_1,\phi_2)=\frac{1}{2} \left\{ (\phi \phi^{*})^6-2(\phi_1^2-\phi_2^2)\left[ (\phi \phi^{*})^2-16 \phi_1^2 \phi_2^2 \right]+1  \right\} $
\item {\bf Vacuum orbit:}

${\cal M}=\frac{D_6}{{\Bbb Z}_2}= \left\{ v^0=1; \, v^1=\frac{1}{2}+i\frac{\sqrt{3}}{2};\, v^2=-\frac{1}{2}+i\frac{\sqrt{3}}{2};\,v^3=-1 ;\,
v^4=-\frac{1}{2}-i\frac{\sqrt{3}}{2};\,
v^5=\frac{1}{2}-i\frac{\sqrt{3}}{2}
\right\} $
\item {\bf Real algebraic curves:}
\[
\phi_1-\frac{\phi_1^7}{7}+3 \phi_1^5 \phi_2^2- 5 \phi_1^3 \phi_2^4+ \phi_1 \phi_2^6
=\gamma;\hspace{0.4cm} \phi_2 - \phi_1^6 \phi_2 +5 \phi_1^4 \phi_2^3 -3\phi_1^2 \phi_2^5+\frac{\phi_2^7}{7} =\gamma_\perp
\]
\item {\bf Kink curves:} there are two choices of $\gamma_\perp$
and three choices of $\gamma$ for which one finds kink curves. The
other kinks associated with the other superpotentials can be
obtained by ${\Bbb Z}_6$ rotations.
\begin{itemize}
\item[a)] $\gamma_\perp =\frac{3 \sqrt{3}}{7}$: kink curve joining $v^1$ with $v^2$

$\gamma_\perp =-\frac{3 \sqrt{3}}{7}$: kink curve joining $v^4$
with $v^5$
\item[b)] $\gamma=\frac{3}{7}$: kink curve joining $v^1$ with $v^5$

$\gamma=-\frac{3}{7}$ : kink curve joining $v^2$ with $v^4$

\item[c)] $\gamma=0$ : kink curve joining $v^0$ with $v^3$
\end{itemize}

\item {\bf Kink energies:} a) $P_0[\phi^{K_{\pm}^1}]=|T|= |W^1(v^{\bar{k}})-W^1(v^{\bar{k}+\bar{1}})|=\frac{6}{7}$

\hspace*{2.6cm} b) $P_0[\phi^{K_{\pm}^2}]=|{\tilde T}|= | W^2(v^{\bar{k}})-W^2(v^{\bar{k}+\bar{2}})|=\frac{6\sqrt{3}}{7}$

\hspace*{2.6cm} c) $P_0[\phi^{K_{\pm}^1}]=|T|= | W^2(v^{\bar{k}})-W^1(v^{\bar{k}+\bar{3}}) |=\frac{12}{7}$

\item {\bf Conserved SUSY charges:} (a) ${\tilde Q}_{\pm}^1 \left|
K_{\pm}^1\right>=0$. (b) and (c) ${\tilde Q}_{\pm}^2 \left|
K_{\pm}^2\right>=0$
\end{itemize}

We now study two even cases.

\item The first and most interesting model occurs for $n=4$. Here, we
find that the kink curves are straight lines in $W$-space (true
for any $n$) and curved in $\phi$-space, in agreement with Reference
\cite{Sa} :

\begin{itemize}
\item {\bf Superpotential:} $W[\phi]=\frac{1}{2}\left(\phi-\frac{\phi^4}{4}\right)$
\[
W^1=\phi_1-\frac{\phi_1^4}{4}+\frac{3}{2} \phi_1^2
\phi_2^2-\frac{\phi_2^4}{4} \hspace{1cm} W^2= \phi_2
\left( 1- \phi_1^3 +\phi_1 \phi_2^2 \right)
\]
\item {\bf Potential:} $U(\phi_1,\phi_2)=\frac{1}{2} \left[ \left(\phi \phi^{*} \right)^3 -2 \phi_1 (\phi_1^2-3 \phi_2^2)+1 \right]
$
\item {\bf Vacuum orbit:} ${\cal M}=\frac{D_3}{{\Bbb Z}_2}=\{v^0=1, v^1=-\frac{1}{2}+i \frac{\sqrt{3}}{2}, v^2= -\frac{1}{2}-i \frac{\sqrt{3}}{2} \} $
\item {\bf Real algebraic curves:}
\[
\phi_1-\frac{\phi_1^4}{4}+\frac{3}{2} \phi_1^2 \phi_2^2-\frac{\phi_2^4}{4}=\gamma ;\hspace{0.9cm} \phi_2 \left( 1- \phi_1^3 +\phi_1 \phi_2^2 \right) =\gamma_\perp
\]
\item {\bf Kink curve:} $\gamma=-\frac{3}{8}$

\item {\bf Kink form factor:} on the kink curve we find $\phi^{K^2_{\mp}}_1=f^{-1} [\pm (x+a)]$ where
\[
f(\phi_1)=\int \frac{d\phi_1}{ \sqrt{\frac{3}{2}+4 \phi_1 +8
\phi_1^4}\sqrt{3 \phi_1^2-\sqrt{\frac{3}{2}+4 \phi_1 +8
\phi_1^4}}}
\]

\item {\bf Kink energy:} $P_0[\phi^{K_{\pm}^2}]=|{\tilde T}|= | W^2(v^{\bar{k}})-W^2(v^{\bar{k}+\bar{1}}) |=\frac{3 \sqrt{3}}{4}$
\item {\bf Conserved SUSY charge:} ${\tilde Q}_{\pm}^2 \left|
K_{\pm}^2\right>=0$
\end{itemize}

\item $n=6$:

\begin{itemize}
\item {\bf Superpotential:} $W[\phi]=\frac{1}{2}\left(\phi-\frac{\phi^6}{6}\right)$
\begin{eqnarray*}
W^1&=&\phi_1-\frac{\phi_1^6}{6}+\frac{5}{2} \phi_1^4 \phi_2^2-
\frac{5}{2} \phi_1^2 \phi_2^4+\frac{\phi_2^6}{6} \\ W^2&=& \phi_2
\left( 1- \phi_1^5 +\frac{10}{3} \phi_1^3 \phi_2^2 -\phi_1
\phi_2^4\right)
\end{eqnarray*}

\item {\bf Potential:} $U(\phi_1,\phi_2)=\frac{1}{2} \left[ (\phi \phi^{*})^5-2 \phi_1( \phi_1^4+5 \phi_2^4-10 \phi_1^2 \phi_2^2)+1 \right]$
\item {\bf Vacuum orbit:} ${\cal M}=\frac{D_5}{{\Bbb Z}_2}= \{v^0=1, v^1= e^{i \frac{2 \pi}{5}} , v^2= e^{i \frac{4 \pi}{5}} ,v^3=e^{i \frac{6 \pi}{5}} ,v^4=e^{i \frac{8 \pi}{5}} \} $
\item {\bf Real algebraic curves:}
\[
\phi_1-\frac{\phi_1^6}{6}+\frac{5}{2} \phi_1^4 \phi_2^2- \frac{5}{2} \phi_1^2 \phi_2^4+\frac{\phi_2^6}{6} =\gamma ;\hspace{0.2cm} \phi_2 \left( 1- \phi_1^5 +\frac{10}{3} \phi_1^3 \phi_2^2 -\phi_1 \phi_2^4\right)=\gamma_\perp
\]
\item {\bf Kink curves:} there are two values of $\gamma$ giving
kink curves: a) $\gamma=-\frac{5}{24} (1+\sqrt{5})$: kink curve
joining $v^2$ with $v^3$, b) $\gamma=\frac{5}{24} (-1+\sqrt{5})$:
kink curve joining $v^1$ with $v^4$. The other kink curves are
obtained through ${\Bbb Z}_5$ rotations.

\item {\bf Kink energies:}
a) $P_0[\phi^{K_{\pm}^2}]=|{\tilde T}|= | W^2(v^{\bar{k}})-W^2(v^{\bar{k}+\bar{1}}) |=\frac{5}{6} \sqrt{\frac{5-\sqrt{5}}{2}}$

\hspace{2.6cm} b) $P_0[\phi^{K_{\pm}^2}]=|{\tilde T}|= | W^2(v^{\bar{k}})-W^2(v^{\bar{k}+\bar{2}}) |=\frac{5}{6} \sqrt{\frac{5+\sqrt{5}}{2}}$

\item {\bf Conserved SUSY charges:} (a) and (b) ${\tilde
Q}_{\pm}^2 \left| K_{\pm}^2\right>=0$
\end{itemize}
\end{itemize}

\begin{figure}[htbp]
\centerline{\epsfig{file=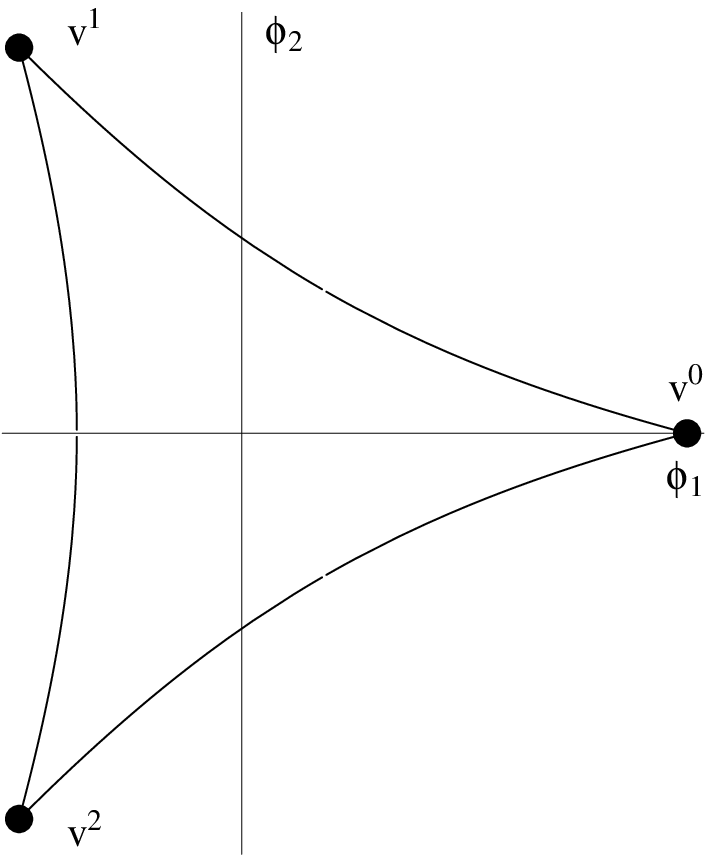,height=3.8cm}
\hspace{0.4cm}\epsfig{file=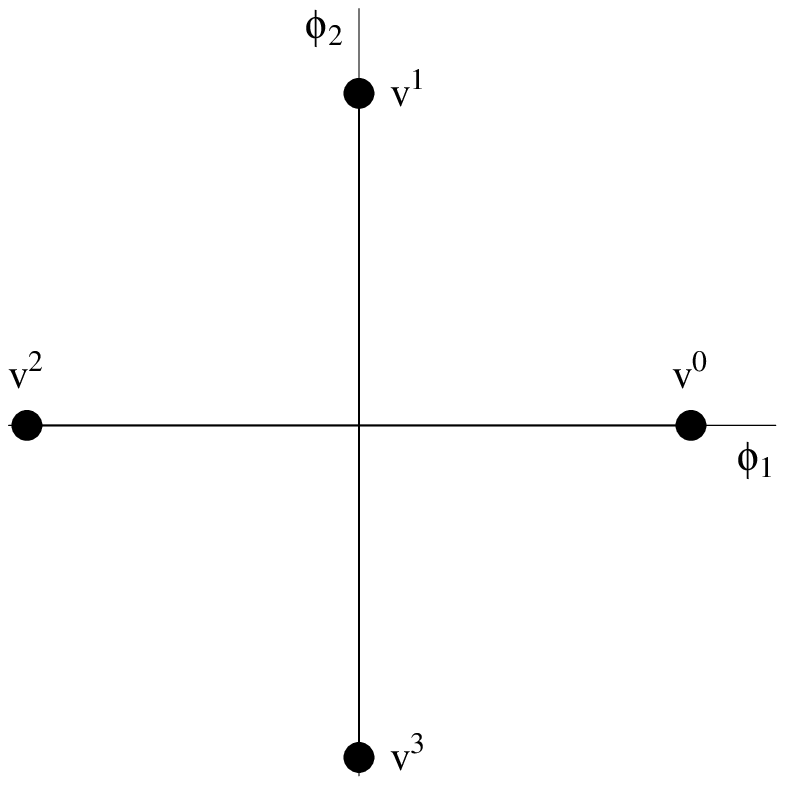,height=3.8cm}\hspace{0.4cm}\epsfig{file=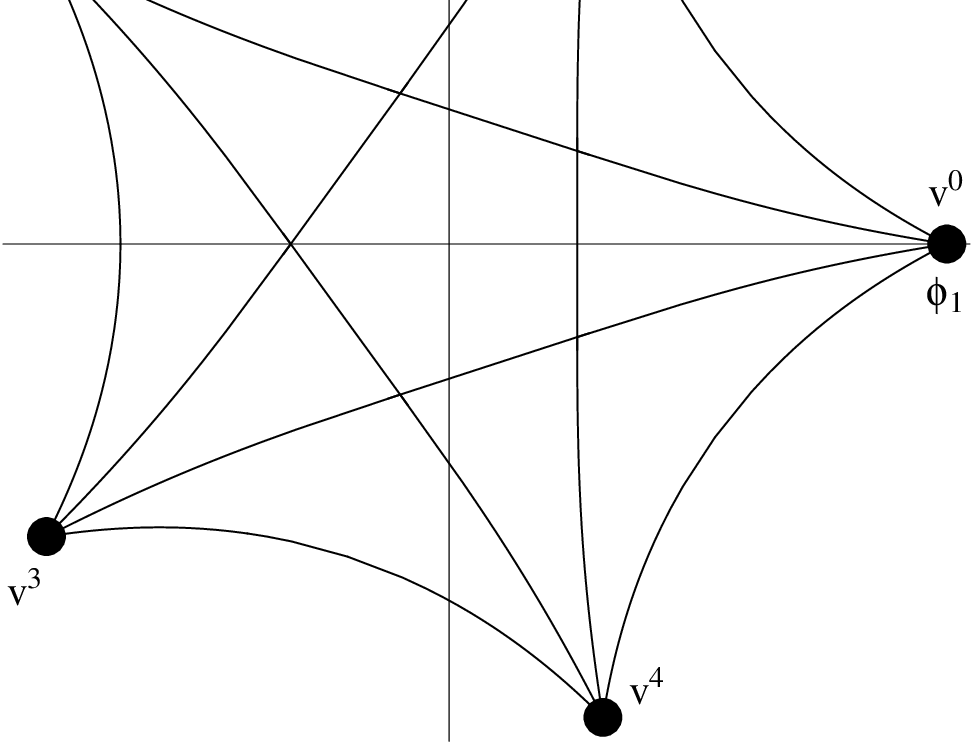,height=3.8cm}\hspace{0.4cm}\epsfig{file=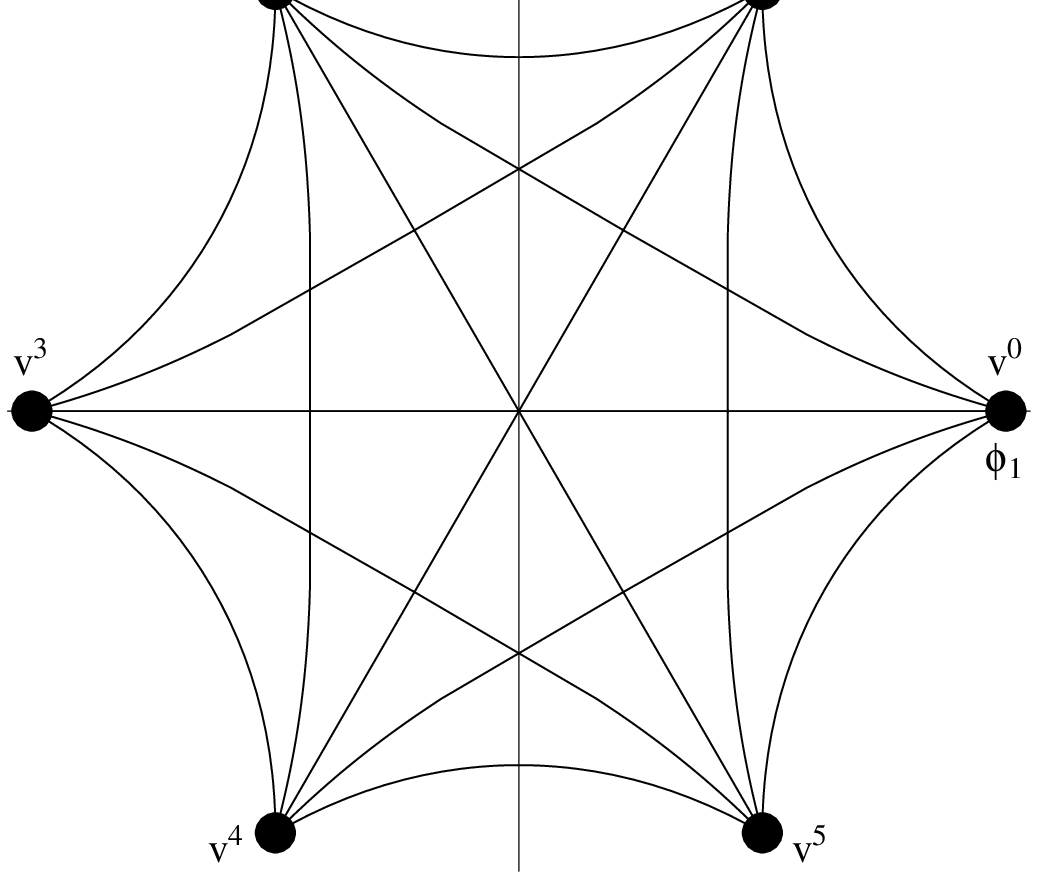,height=3.8cm}
} \caption{\small \it Kink curves in the $n=4$, $n=5$, $n=6$ and
$n=7$ models }
\end{figure}


\begin{thebibliography}{99}
\addcontentsline{toc}{section}{References}

\bibitem{Sh} M. Shifman, A. Vainshtein and M. Voloshin, Phys.Rev. ${\bf D59}$ (1999) 45016.

\bibitem{Gi} G. Gibbons and P. Townsend, Phys. Rev. Lett. {\bf 83} (1999) 1727.

\bibitem{Fr} P. Freund, {\sl \lq\lq Introduction to Supersymmetry"}, Cambridge University Press, 1986, New York.

\bibitem{OW} E. Witten, D. Olive, Phys. Lett. {\bf 78B}  (1978) 97

\bibitem{Ba} D. Bazeia and F. Brito, Phys. Rev. Lett. {\bf 84}
(2000) 1094

\bibitem{Va} S. Cecotti and C. Vafa, Com. Math. Phys. {\bf 158}
(1993) 569

\bibitem{To} P. Townsend, {\sl \lq\lq Three Lectures on Supersymmetry and Extended Objects"}, in NATO ASI Series C: Vol. 409, edited by L. Ibort and M.A. Rodriguez, Kluwer Academic Publisher, Dordrech, 1993.

\bibitem{Sa} P. Saffin, Phys. Rev. Lett. {\bf 83} (1999) 4249 ; S. Carroll, S. Hellerman and M. Trodden,
hep-th/9905217, Phys. Rev. D, to appear; D. Binosi and T. ter
Veldhuis, hep-th/9912081, Phis. Lett. B, to appear.
\end{thebibliography}
\end{document}